\documentclass[11pt,oneside,a4paper]{article}
\usepackage{graphicx}
\usepackage{lipsum,hyperref}
\usepackage{times,authblk}
\title{Relativistic model for cold spherical interstellar gas clouds}
\author{D\'{a}niel Barta \\ {\footnotesize email: \href{mailto:barta.daniel@wigner.mta.hu}{barta.daniel@wigner.mta.hu}}}
\affil{%
	{\footnotesize Institute for Particle and Nuclear Physics, Wigner Research Centre for Physics, Hungarian Academy of Sciences,
	Konkoly-Thege Mikl\'{o}s \'{u}t 29-33., H-1121 Budapest, Hungary}
}%

\begin{document}

\maketitle

\abstract{We investigate insterstellar gas spheres by determining the metric functions, the material distribution,
  and the features of particle orbits in terms of stability and geodesics. An exact solution of the Einstein's
  equations for interstellar gas clouds is derived that is compatible with the results of recent astronomical
  measurements. The solution determines the distribution of pressure and density, and it is suitable to
  describe the energy, speed, trajectory, and further relevant physical features of the cloud's particles.
  We describe the spacetime inside the nebula and give the density profile and the geodesics of particles.
  We find that circular orbits are stable and the cloud rotates rigidly by an angular velocity that is
  inversely proportional to the radius.}

\section{Introduction}

The general relativistic gravitational fields are described by Einstein's equations. Although due to their non-linearity one unawares 
encounter difficulties in solving them for spherically symmetric static gas cloud. \textsl{\cite{burlankov}} and \textsl{\cite{goldman}} has studied perfect fluid 
spacetimes and introduced generating functions by choices of new variables in order to replace the second order ODEs with algebraic equations. 
For retain generality, \textsl{\cite{fodor}} used no equation of state either.\\
By specializing the problem for ideal gas, one may make use of the linear relation between pressure and density. One may also characterize 
the matter by the stress-energy tensor of perfect fluid, while keeping in mind the fact that the gasous medium is 
compressible contrary to the perfect fluids. In addition, one should impose a few physical criteria for the maximal mass of the 
nebula, boundary condition on the pressure and density, as well as for the sound of speed in the medium.\\
Evidently, the corresponding quantities interpreted in the context of general relativity must be identical with the results 
of \textsl{\cite{bohigas}} and \textsl{\cite{kritsuk}} in the classical limit. \textsl{\cite{hobson}} lend assistance to calculate the geodetics and 
further derived quantities.


\section{Basic properties of a self-gravitating spherically symmetric static cloud}

Consider an isolated interstellar nebula remote from any other matter, and assume that the hydrostatic pressure is balanced by
the cloud's self-gravitation. The mass within a distance $r$ from the centre of the cloud is given by
\begin{equation} \label{felhotomeg}
M_{r} = 4\pi\int_{0}^{r} r'^{2}\rho(r') \mathrm{d}r'.
\end{equation}
The most dense and heavy of all nebulae are the giant molecular clouds composed by mostly gas and some dust. 
For the sake of simplicity, assume that the considered medium consists only of cold neutral gas. In this case collisions between 
these low-energy particles are rare and weak, and have no significant effect on the system. Hence the material of the 
cloud can be realistically regarded as ideal gas. Since temperature is nearly constant ($10-20$ K for molecular clouds), 
equation of state becomes
\begin{equation} \label{izotermikus}
p = c_{s}^{2}\rho,
\end{equation}
a linear relation of pressure $p$ and density $\rho$, the coefficient $c_{s}$ is the isothermal sound speed within the gaseous medium.
On the basis of the isothermal equation of state (\ref{izotermikus}) and (\ref{felhotomeg}), the total mass of the cloud $M_{\mathrm{R}}$ is 
expressed by the average pressure $\bar{p}$ (see the following Table \ref{molekulafelho_adatok}.) as $M_{\mathrm{R}} = 4\pi R^{3}\bar{p}/3c_{s}^{2}$. 
By the comparsion of these with a given radial pressure distribution $p(r)$, the value of average pressure can be written as
\begin{equation} \label{atlagosnyomas}
\bar{p} = \frac{R^{3}}{3}\int^{R}_{0} p(r)r^{2}\mathrm{d}r.
\end{equation}
It is important to state that the following criteria must be met for real physical systems:
\begin{enumerate}
 \item $M_{\mathrm{R}} \leq M_{\mathrm{BE}}$ where $M_{\mathrm{BE}}$ refers to the Bonnor--Ebert mass given by $M_{\mathrm{BE}} = c_{\mathrm{BE}}c_{s}^{4}p^{-1/2}$,
 where $c_{\mathrm{BE}} \simeq 1.18$ is a dimensionless constant (see in \textsl{\cite{girichidis}}). This is the largest mass that an isothermal gas sphere embedded in a 
 pressurized medium can have while still remaining in hydrostatic equilibrium.
 \item $\rho(r)$, $p(r) > 0$ and $\mathrm{d}\rho/\mathrm{d}r$, $\mathrm{d}p/\mathrm{d}r < 0$ everywhere in the cloud, the maximum of the density and pressure are 
 $\rho_{0}$ and $p_{0}$ in $r = 0$. On the border of the nebula, the density and pressure distributions must satisfy the boundary conditions
 {
 \begin{equation} \label{felho-hatarfeltetelek}
 \displaystyle \lim_{r \to R}\rho(r) =  \lim_{r \to R}p(r) = \lim_{r \to  R}\frac{\mathrm{d}\rho}{\mathrm{d}r} = \lim_{r \to  R}\frac{\mathrm{d}p}{\mathrm{d}r} = 0.
 \end{equation}}
 These conditions can be easily justified: the first one expresses the simple fact that the density and pressure
 disappears, the second one says that they do not change on the border of the cloud. So the matter does not suddenly vanish
 on the border, but steadily aligns into the environment.
 \item The speed of sound $c_{s}$ in the medium must be less than the speed of light, that is
 \begin{equation} \label{hangsebesseg}
 c_{s}^{2} = \frac{\mathrm{d}p}{\mathrm{d}\rho} < 1.
 \end{equation}
\end{enumerate}
   \begin{table}[ht]
      \caption[]{Physical properties of a typical cold giant molecular cloud in SI according to \textsl{\cite{ferriere}}}
      \label{molekulafelho_adatok}
     $$ 
         \begin{array}{p{0.5\linewidth}l}
            \hline
            \noalign{\smallskip}
            Composition      &  $neutral H$_{2} \\
            \noalign{\smallskip}
            Radius & R = 70 $ pc$ \\
            \noalign{\smallskip}
            Constant temperature & T = 10 $ K$ \\
            \noalign{\smallskip}
            Average density & \bar{\rho} = 3.3475 \times 10^{-15} $ kg$/$m$^{3} \\
            \noalign{\smallskip}
            Average pressure & \bar{p} = 4.5827 \times 10^{-16} $ Pa$ \\
            \noalign{\smallskip}
            Total mass$^{1}$ & M = 0.5433 \times 10^{23} $ $M_{\odot} \\
            \noalign{\smallskip}
            \hline
         \end{array}
     $$ 
      \begin{list}{}{}
         \item[$^{1}$] $M_{\odot}=1.9891 \times 10^{30}\,\mathrm{kg}$ is the Sun's mass
      \end{list}
   \end{table}


\section{Field equations for the compact gas cloud}

The metrics of a general stationary spherically symmetric configuration can be written in area coordinates as              
\begin{equation} \label{sch-ivelem}
ds^{2} = -e^{\nu}\mathrm{d}t^{2} + e^{\lambda}\mathrm{d}r^{2} + r^{2}(d\vartheta^{2} + \sin^{2}\vartheta \mathrm{d}\varphi^{2})
\end{equation}                      
where $\nu$ and $\lambda$ are functions of the radial coordinate $r$. Consider the above described spacetime region is 
filled with ideal gas, $u^{\nu}$ is the contravariant velocity vector of gas particles. The stress-energy tensor is 
equivalent with the stress-energy tensor $T_{\mu\nu} = \displaystyle \left(\rho + p\right)u_{\mu}u_{\nu} + pg_{\mu\nu}$ 
for perfect fluid. Calculating the Einstein's equation $G_{\mu\nu} = 8\pi T_{\mu\nu}$, one obtains the mass density, 
the radial and the angular directional pressure as
{
\begin{equation} \label{folyad-schwarzschild}
\begin{array}{lcl}
8\pi r^{2}\rho & = & \displaystyle e^{-\lambda}(r\lambda' - 1) + 1 \\
8\pi r^{2}p_{r} & = & \displaystyle -e^{-\lambda}(r\nu'+ 1) + 1 \\
32\pi rp_{\vartheta} & = & \displaystyle -e^{-\lambda}(2r\nu'' - r\lambda'\nu' + r\nu'^{2} + 2\nu' - 2\lambda')
\end{array}
\end{equation}}
\hspace{-18pt} where the prime denotates derivates with respect to the radial coordinate $r$. \textsl{\cite{fodor}} has shown that 
the set of differential equations (\ref{folyad-schwarzschild}) can be reduced to algebraic ones with integration required 
only for one metric function but not the physical variables $\rho$ and $p$. From this point on we slightly modify Fodor's method 
and apply it to isothermal ideal gas.

Due to the isotropic configuration, $p \equiv p_{r} = p_{\vartheta}$ implies one can 
require one more field equation
{
\begin{equation} \label{teregyenlet}
r(r\nu' + 2)\frac{\mathrm{d}}{\mathrm{d}r}e^{-\lambda} + (2r^{2}\nu'' + r^{2}\nu'^{2} - r\nu' - 4)e^{-\lambda} + 4 = 0
\end{equation}}
\hspace{-18pt} by extracting the last equation from the second one. Regarding the coefficient of $\mathrm{d}e^{-\lambda}/\mathrm{d}r$, it turns out to be practical to 
introduce a pair of new variables
\begin{equation} \label{beta-definicio}
\alpha = -\lambda'e^{-\lambda}\beta^{2} \hspace{10pt} , \hspace{10pt} \beta = \displaystyle \frac{r\nu'}{2} + 1.
\end{equation}
Then the field equation (\ref{teregyenlet}) reduces to a second order algebric equation in $\beta$, namely
\begin{equation} \label{alpha-masodfokuegyenlet}
2(\alpha + 1)\beta^{2} + (r\alpha' + 8\alpha)\beta + 4\alpha = 0.
\end{equation}
For any function $\alpha$ the quadratic equation (\ref{alpha-masodfokuegyenlet})
is solved by the real roots
{
\begin{equation} \label{beta-megoldas}
\beta_{\pm} = \frac{8\alpha - r\alpha' \pm \sqrt{(r\alpha' + 8\alpha)^{2} - 32\alpha(\alpha + 1)}}{4(\alpha + 1)}
\end{equation}}
\hspace{-18pt} where the discriminant must be non-negative. The only physically relevant solution as \textsl{\cite{fodor}} has already shown is $\beta_{+}$, 
since its non-positive counterpart always belongs to a non-positive, hence non-physical mass density. 
The metric functions belonging to $\beta$ are formally given by the definitions (\ref{beta-definicio}) as
\begin{equation} \label{nu-lambda-integral}
\lambda = \displaystyle \ln\left(\frac{\beta^{2}}{\alpha}\right) \hspace{7pt} , \hspace{7pt} \nu = \displaystyle \int_{0}^{r}\frac{2(\beta - 1)}{r}\mathrm{d}r + \nu_{0}
\end{equation}
where the constant $\nu_{0}$ determines the scaling of the time coordinate $t$. One can also calculate the pressure and density
\begin{equation} \label{suruseg-alphaval}
\rho = \frac{1 - (r\alpha/\beta^{2})'}{8\pi r^{2}} \hspace{7pt} , \hspace{7pt} p = \displaystyle \frac{(2\beta - 1)\alpha - \beta^{2}}{8\pi \beta^{2}r^{2}}
\end{equation}
by substituting functions $\alpha$ and $\beta$ into the first two field equations of Eq. (\ref{folyad-schwarzschild}). The simple, but 
still realistic choice for the generating function $\alpha$ is the ratio of two polynomials of the radial coordinate $r$. The lowest 
degree form which is physically valid for a compact fluid or gaseous sphere is
\begin{equation} \label{valasztott-alpha}
\alpha = 1 + \frac{A^{2}r^{2}}{1 + Br^{2}}
\end{equation}
where $A$ and $B$ are positive constants associated by inverse first and second power of distance dimensions. It is advisable to 
introduce a further new non-negative real constant $C^{2} = 2B/A^{2} - 2$ and use it in place of constant $B$. 
In order to eliminate the square root appeared in Eq. (\ref{beta-megoldas}) while expressing $\beta$, a new radial 
variable defined by
\begin{equation} \label{sugar-transzf}
\sinh\xi = 2C\frac{1 + Br^{2}}{3 + 4Br^{2}}
\end{equation}
will be introduced. Then the centre gets into $\xi_{c} = $ arcsinh$(2C/3)$, and the spatial infinity 
$\xi_{\infty} = $ arcsinh$(C/2)$, and the new variable is restricted by $0 < \xi_{\infty} \leq \xi \leq \xi_{c}$. 
Through (\ref{valasztott-alpha}) the generating functions $\alpha$ and $\beta$ become
{
\begin{equation}
\alpha = \displaystyle\frac{(C^{2} - 4)\sinh\xi + 4C}{(C^{2} + 2)\sinh\xi}, \hspace{5pt}
\beta = \displaystyle\frac{C\coth(\xi/2) - 2}{1 + C\tanh(\xi/2)}.
\end{equation}}
The equations (\ref{nu-lambda-integral}) provide the metric functions\footnote{The metric functions as well as the density 
and pressure are identical to Fodor's results (61)-(63) and (65)-(66).} and the inner Schwarzschild metrics appears to be
\[
\mathrm{d}s^{2} = e^{\lambda}\frac{C^{2}\cosh^{2}\xi \mathrm{d}\xi^{2}}{4A^{2}(C^{2} + 2)(2C - 3\sinh\xi)(2\sinh\xi - C)^{3}}
\]
\[
\hspace{7pt} -e^{\nu}\mathrm{d}t^{2} + \frac{2C - 3\sinh\xi}{A^{2}(C^{2} + 2)(2\sinh\xi - C)}(\mathrm{d}\vartheta^{2} + \sin^{2}\vartheta \mathrm{d}\varphi^{2})
\]
\hspace{-18pt} by using the charasteric $\xi$ as radial coordinate. The constant $A$ corresponds to a constant conformal transformation of 
the metrics. From Eq. (\ref{folyad-schwarzschild}) both the density and the pressure are expressable by a ratio of two polynomials 
of hyperbolic function of the radial coordinate $\xi$. 

\begin{figure}[ht!]
\centering
\includegraphics[scale=0.3]{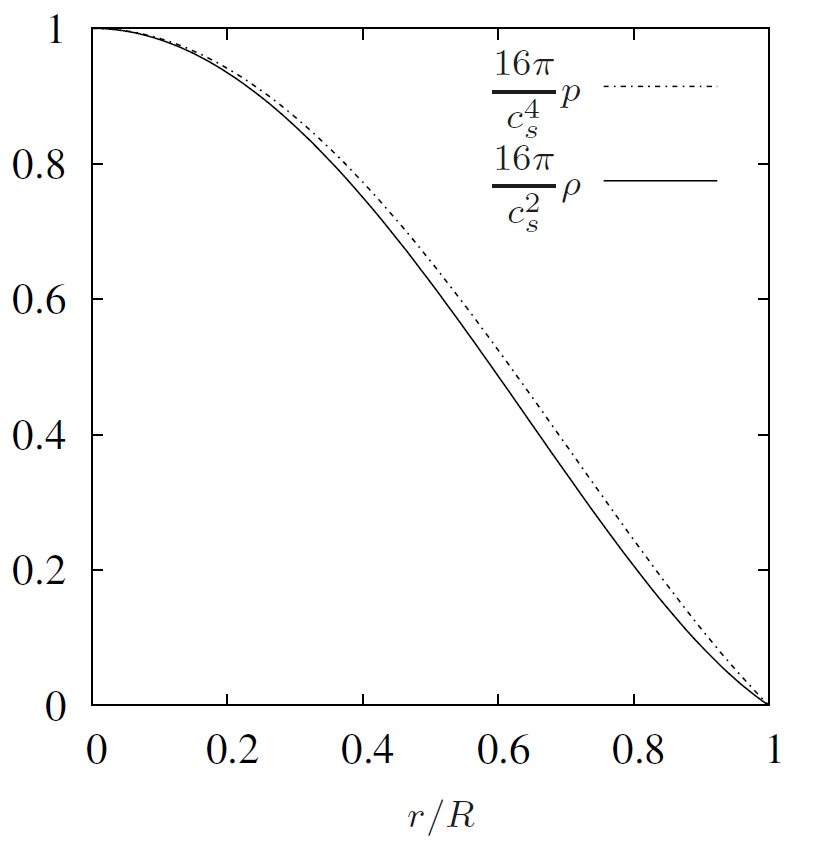}
\caption{The pressure and density profile in the interval $0 < r/R < 1$, normalized to the central values $p_{c}$ and $\rho_{c}$.}
         \label{FigProfile}%
\end{figure}

Moving away from the centre of the cloud due to the conditions (\ref{felho-hatarfeltetelek}), the pressure monotonously tends to 
zero at $r = R$, on the border of the cloud. Any choice of constants $A$ and $C$ satisfies the restriction (\ref{hangsebesseg}) on 
the speed of sound in the medium. One takes $C^{2} = 2B/A^{2} - 2$ into consideration and assumes $B \ll 1 \ll C$, then 
$p$ and $\rho$ vanish simultaneously at $r = R$ if and only if $B = 4/R^{2}$. This restriction implies $A = 8c_{s}^{2}/R$. 
By eliminating the variable $\xi$ via the transformation (\ref{sugar-transzf}), one can formulate the functions of state
{
\begin{equation} \label{sugarfuggonyomas}
\rho = \displaystyle \frac{8Br^{2} - 3}{4\pi C (4Br^{2} - 1)^{-1}}, \hspace{10pt} p = \displaystyle \frac{2Br^{2} - 1}{4\pi C^{2}(4Br^{2} - 1)^{-1}}
\end{equation}}
\hspace{-18pt} in terms of polynomials of the natural radial coordinate $r$, see Fig. \ref{FigProfile}. As it was required, if $B \ll 1 \ll C$ 
then the equation of state is nearly linear for every $r \leq R$, therefore
\[
\frac{p}{\rho} = \frac{1}{C}\frac{2Br^{2} - 1}{8Br^{2} - 3} = c_{s}^{2} < 1
\]
fixes the last constant as $C = 4/c_{s}^{2}$. Consequently, they differ from one another only by a 
constant factor, thus verifying the legitimacy of the isothermal equation of state (\ref{izotermikus}).
The obtained density and pressure distribution correspond with the classical results of \textsl{\cite{bohigas}} (see his Fig. 6.) and 
\textsl{\cite{kritsuk}}, hence the metric functions consistent with the distributions must be valid.

Similarly, the metric functions $\nu$ and $\lambda$ can be expressed as functions of the radial coordinate $r$. The line element
{
\begin{equation} \label{sch-ivelem2}
\begin{array}{rcl}
\mathrm{d}s^{2} = \displaystyle -\frac{c_{s}^{2}}{4}\left(1 + \frac{c_{s}^{2}}{4}\frac{r^{2}}{R^{2}}\right)\mathrm{d}t^{2} + 
\exp\left(-\frac{c_{s}^{2}}{2}\frac{r^{2}}{R^{2}}\right)\mathrm{d}r^{2} \\
+ \displaystyle r^{2}(\mathrm{d}\vartheta^{2} + \sin^{2}\vartheta \mathrm{d}\varphi^{2})
\end{array}
\end{equation}}
\hspace{-18pt} is immediately obtaind by the insertion of the metric functions into the general form of 
Schwarzschild metrics (\ref{sch-ivelem}).

\begin{figure}[ht]
\centering
\includegraphics[scale=0.3]{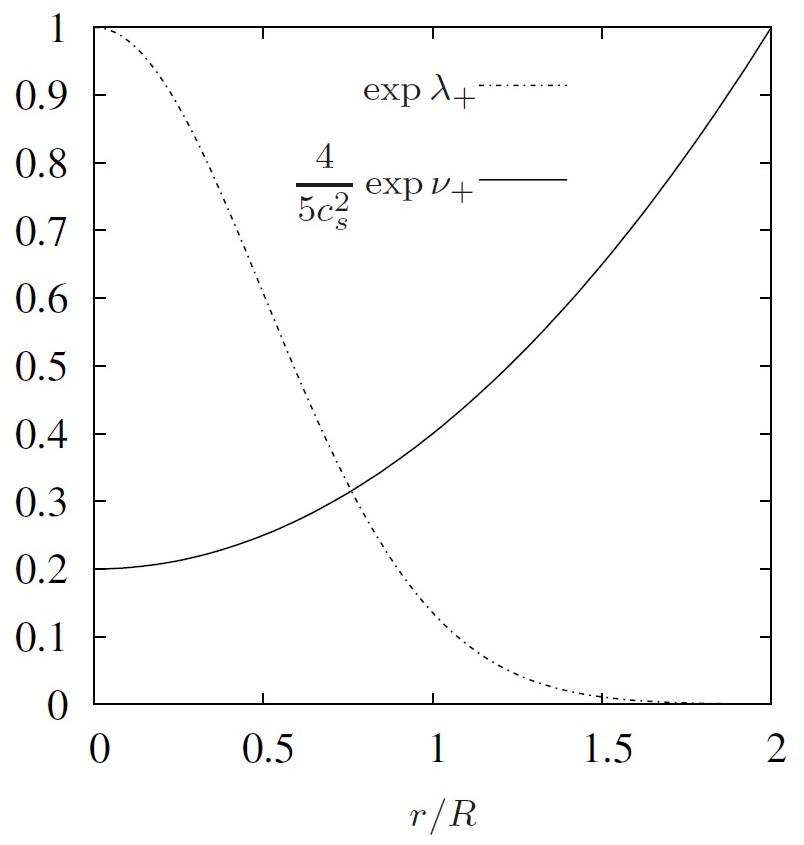}
\caption{The evolution of the normalized metric functions $\exp\lambda_{+}$ and $\exp\nu_{+}$ within a fraction of distance $R$.}
         \label{FigMetrics}%
\end{figure}


\section{Lagrangian function and geodetics of the particles}

In the following sections the geodetics in the Schwarzschild geometry in several textbooks, like \textsl{\cite{hobson}} originally studied 
the behaviour of massive particles and photons will be briefly reviewed according to our special geometry.

For the Schwarzschild metrics (\ref{sch-ivelem2}) the relativistic Lagrangian function 
$L = g_{\mu\nu}\dot{x}^{\mu}\dot{x}^{\nu}$ of the particles in the investigated nebula is 
\begin{equation}
L = -e^{\nu}\dot{t}^{2} + e^{\lambda}\dot{r}^{2} + r^{2}(\dot{\vartheta}^{2} + \sin^{2}\vartheta \dot{\varphi}^{2})
\end{equation}
where the dot denotates derivates with respect to the proper time coordinate $\tau$. By substituting this form for $L$ into 
the Euler--Lagrange equations $\displaystyle \frac{\mathrm{d}}{\mathrm{d}\tau}\left(\frac{\partial L}{\partial \dot{x}^{\mu}}\right) - \frac{\partial L}{\partial x^{\mu}} = 0$,
result the geodetic equations. Since the eqation for $\mu = 3$ is satisfied by $\vartheta = \pi/2$, it is sufficient to keep only 
the set of three equations independent of $\vartheta$:
\begin{equation} \label{geodetikusok}
\begin{array}{rcl}
e^{\nu}\dot{t} = L_{t} \\
\displaystyle \ddot{r} + \frac{1}{2}\frac{\mathrm{d}\lambda}{\mathrm{d}r}\dot{r}^{2} + \frac{1}{2}\frac{\mathrm{d}\nu}{\mathrm{d}r}e^{\nu-\lambda}\dot{t}^{2} - 
re^{-\lambda}\dot{\varphi}^{2} = 0 \\
r^{2}\dot{\varphi} = L_{\varphi}
\end{array}
\end{equation}
The two simplest equations are derived immediately since the Lagrangian is not an explicit function of $t$ or $\varphi$. 
The appearing constants $L_{t}$ and $L_{\varphi}$ proportional to the total energy and the angular momentum of the particles. 
It is expedient to replace the complicated second equation of Eq. (\ref{geodetikusok}) by the first integral 
$g_{\mu\nu}\dot{x}^{\mu}\dot{x}^{\nu} = -1$ of the geodetic equation, since the worldline of a massive partice is timelike. In 
our case, it takes the form
\begin{equation} \label{r-geodetikus}
-e^{\nu}\dot{t}^{2} + e^{\lambda}\dot{r}^{2} + \frac{1}{2}r^{2}\dot{\varphi}^{2} = -1.
\end{equation}
By substituting the two original expressions of (\ref{geodetikusok}) into (\ref{r-geodetikus}), one obtain the combined energy equation
\begin{equation} \label{energia-egyenlet}
\dot{r}^{2} + \frac{L_{\varphi}^{2}}{r^{2}}e^{-\lambda} = \left(L_{t}^{2}e^{-\nu} - 1\right)e^{-\lambda}
\end{equation}
for the radial coordinate valid inside the cloud. Outside of the cloud the customary equation
$\dot{r}^{2} + (1 - 2M/r)L_{\varphi}^{2}/r^{2} - 2M/r = (L_{t}^{2} - 1)$
governs the motion of particles. Note that the right-hand side is a constant of motion, $L_{t} \propto E$ as previously stated. 
The constant of proportionality is fixed by requiring $E = m_{0}$ for a particle at rest at $r = \infty$, $m_{0}$ is the mass of the 
particle at rest. Letting $r \to \infty$ and $\dot{r} = 0$ in the equation, $L_{t}^{2} = 1$ thus is required. Hence, one must has 
$L_{t} = E/m_{0}$ where $E$ is the total energy of the particle in its orbit.

The shape of a particle orbit is given by using the last equation of Eq. (\ref{geodetikusok}) to express $\dot{r}$ in the 
(\ref{energia-egyenlet}) as
\begin{equation} \label{dotr}
\frac{\mathrm{d}r}{\mathrm{d}\tau} = \frac{\mathrm{d}r}{\mathrm{d}\varphi}\frac{\mathrm{d}\varphi}{\mathrm{d}\tau} = \frac{L_{\varphi}}{r^{2}}\frac{\mathrm{d}r}{\mathrm{d}\varphi}.
\end{equation}
Furthermore, if one parametrizes Eq. (\ref{r-geodetikus}) by $\tilde{r} \equiv 1/r$, one obtains
$\displaystyle \left(\frac{\mathrm{d}\tilde{r}}{\mathrm{d}\varphi}\right)^{2} + \tilde{r}^{2}e^{-\lambda} = \frac{1}{L_{\varphi}^{2}}\left(L_{t}^{2}e^{-\nu} 
- 1\right)e^{-\lambda}$. Finally, the differentiation with respect to $\varphi$ provides the orbits
{
\begin{equation} \label{reszecskepalya}
\frac{\mathrm{d}^{2}\tilde{r}}{\mathrm{d}\varphi^{2}} + \tilde{r}e^{-\lambda} = \frac{L_{t}^{2}(\nu' 
+ \lambda')e^{-\nu} - \lambda'}{2L_{\varphi}^{2}\tilde{r}^{2}}e^{-\lambda} - \frac{\lambda'}{2}e^{-\lambda}
\end{equation}}
\hspace{-18pt} for a particle in the equatorial plane $\vartheta = \pi/2$ where the prime denotes derivates with respect to the radial 
coordinate $r$. The particle orbits have two special cases, namely the radial motion where $\dot{\varphi} = 0$ and the 
circular motion where $\dot{r} = 0$. Since the density of the nebula is constant in time, one might ignore the radial motion 
and focus on investigating the latter motion.


\section{Circular motion on bounded and stabil orbits, velocity of the gas particles}

In the equatorial plain for circular motion, one has $r = $ constant, and thus $\dot{r} = \ddot{r} = 0$. This restriction in 
accordance of (\ref{dotr}) imposes $\tilde{r}' = \dot{\tilde{r}}/L_{\varphi}\tilde{r}^{2} = 0$; consequently  $\tilde{r}''$ 
is zero too. Setting $\tilde{r} = 1/r = $ constant in the equation of orbits (\ref{reszecskepalya}), one has
$L_{\varphi}^{2} = \displaystyle \frac{1}{2}r^{3}L_{t}^{2}(\nu' + \lambda')\nu'e^{-\nu}$.
Beside replacing the differentials of the metrics functions of Eq. (\ref{sch-ivelem2})
\begin{equation} \label{nu-lambda-derivaltak}
\frac{\mathrm{d}\lambda}{\mathrm{d}r} = -\frac{c_{s}^{2}r}{R^{2}} \hspace{10pt} , \hspace{10pt} \displaystyle 
\frac{\mathrm{d}\nu}{\mathrm{d}r} = \frac{c_{s}^{2}r}{2R^{2}}\left(1 + \frac{c_{s}^{2}}{4}\frac{r^{2}}{R^{2}}\right)^{-1}
\end{equation}
\hspace{-18pt} in the energy equation (\ref{energia-egyenlet}) in addition to the condition $\dot{r} = 0$, one can identify the constants of 
motion as
\begin{equation} \label{k-h-osszefugges}
L_{t} = \frac{c_{s}}{2}\left(1 + \frac{c_{s}^{2}r^{2}}{4R^{2}}\right) \hspace{10pt} , \hspace{10pt} 
\displaystyle L_{\varphi} = \frac{c_{s}}{2}\frac{r^{2}}{R}.
\end{equation}
It has been shown that $E = L_{t}m_{0}$ is the total energy of a particle of rest mass $m_{0}$ in a circular of radius $r$. Subsequently 
one can circumscribe the bounded orbits by requiring $E < m_{0}$, so as long as $L_{t} = 1$. The limits on $r$ for the orbit to be 
bound is given by $\displaystyle 1 = \frac{c_{s}}{2}\left(1 + \frac{c_{s}^{2}r^{2}}{4R^{2}}\right)$ which is satisfied when
\begin{equation}
r = \frac{2R}{c_{s}}\sqrt{\frac{2}{c_{s}} - 1}.
\end{equation}
The first and third geodetic equations in Eq. (\ref{geodetikusok}) immediately shows that the components of 4-velocity of a particle 
are simply
\begin{equation}
[u^{\mu}] = \left[\frac{2}{c_{s}}, 0, 0, c_{s}\frac{r}{2R}\sin\vartheta\right]
\end{equation}
in the coordinate system $(t,r,\vartheta,\varphi)$. The geodetic equations specify the circular trajectory $\varphi(\tau)$ and
the orbital period $T = 2\pi/\dot{\varphi}$, which according to (\ref{k-h-osszefugges}) is $T = 4\pi R/c_{s}$
by substituting $L_{\varphi}$ from (\ref{k-h-osszefugges}). Although $r$ is not the radius of the orbit, it is readily conceivable 
that the spatial distance travelled in one complete revolution is $2\pi r$, just as in the Newtonian case. Instead of parametrizing 
$\varphi$ in the proper time, one can alternatively describe it by $\mathrm{d}\varphi/\mathrm{d}t = c_{s}^{2}/4R$ in terms of 
coordinate time $t$, thus the 4-velocity is
\begin{equation}
[u^{\mu}] = \left[1, 0, 0, \frac{c_{s}^{2}}{4}\frac{r}{R}\sin\vartheta\right].
\end{equation}


\section{Conclusions}
        For cold isotermal molecular clouds in gravitational and thermal equilibrium, 
        the pressure correlates linearily to the density. Their distribution can expressed by 
        a decreasing function of radius in terms of only the speed of sound in the medium and 
        the size of the cloud. The profils correspond with astrophysical measurments.
        The metric functions provides us Lagrangians that determines the geodetics of the 
        particles; all the circular orbits are stable, thus the cloud rotate rigidly and theoretically
        it remains stable permanently.
        The value of the four-velocity of a particle slightly differs from the one observed in an ordinary Schwarzschild 
        spacetime, but the angular velocity is inversely proportional to the radius.

\end{document}